\documentclass[fleqn,usenatbib]{mnras}

\usepackage{mathptmx}
\usepackage[T1]{fontenc}

\DeclareRobustCommand{\VAN}[3]{#2}
\let\VANthebibliography\thebibliography
\def\thebibliography{\DeclareRobustCommand{\VAN}[3]{##3}\VANthebibliography}

\usepackage{graphicx}	% Including figure files
\usepackage{amsmath}	% Advanced maths commands
\usepackage{amssymb}	% Extra maths symbols
\defcitealias{Sebastian24b}{S24}
\defcitealias{Bender2012}{B12}
\defcitealias{triaud22}{T22}
\title[Dynamical Masses of Kepler-16~AB]{EBLM XV -- Revised dynamical masses for the circumbinary planet host Kepler-16 AB, using the SOPHIE spectrograph}

\author[D. Sebastian et al.]{
D. Sebastian$^{1}$ $^{\href{https://orcid.org/0000-0002-2214-9258}{\includegraphics[scale=0.5]{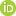}}}$\thanks{E-mail: D.Sebastian.1@bham.ac.uk}, 
I. Boisse$^{2}$, 
A. Santerne,$^{2}$ $^{\href{https://orcid.org/0000-0002-3586-1316}{\includegraphics[scale=0.5]{figures/orcid.jpg}}}$,
A. H.M.J. Triaud$^{1}$ $^{\href{https://orcid.org/0000-0002-5510-8751}{\includegraphics[scale=0.5]{figures/orcid.jpg}}}$,
T.\,A. Baycroft$^{1}$ $^{\href{https://orcid.org/0000-0002-3300-3449}{\includegraphics[scale=0.5]{figures/orcid.jpg}}}$, 
Y.\,T. Davis$^{1}$ $^{\href{https://orcid.org/0009-0000-6625-137X}{\includegraphics[scale=0.5]{figures/orcid.jpg}}}$, 
\newauthor
M. Deleuil$^{2}$ $^{\href{https://orcid.org/0000-0001-6036-0225}{\includegraphics[scale=0.5]{figures/orcid.jpg}}}$, 
S. Grouffal$^{2}$, 
G. H\'ebrard$^{3,4}$,
N. Heidari$^{3}$ $^{\href{https://orcid.org/0000-0002-2370-0187}{\includegraphics[scale=0.5]{figures/orcid.jpg}}}$, 
D. V. Martin$^{5}$ $^{\href{https://orcid.org/0000-0002-7595-6360}{\includegraphics[scale=0.5]{figures/orcid.jpg}}}$, 
P.\,F.L. Maxted$^{6}$ $^{\href{https://orcid.org/0000-0003-3794-1317}{\includegraphics[scale=0.5]{figures/orcid.jpg}}}$, 
R.\,P. Nelson$^{7}$, 
\newauthor
Lalitha S.$^{1, 8}$ $^{\href{https://orcid.org/0000-0001-8102-3033}{\includegraphics[scale=0.5]{figures/orcid.jpg}}}$,
M. G. Scott$^{1}$
$^{\href{https://orcid.org/0009-0006-3846-4558}{\includegraphics[scale=0.5]{figures/orcid.jpg}}}$,
O.\,J. Scutt$^{1}$ $^{\href{https://orcid.org/0000-0002-7603-3525}{\includegraphics[scale=0.5]{figures/orcid.jpg}}}$,
M. Standing$^{9}$ $^{\href{https://orcid.org/0000-0002-7608-8905}{\includegraphics[scale=0.5]{figures/orcid.jpg}}}$, 
\\
$^{1}$ School of Physics \& Astronomy, University of Birmingham, Edgbaston, Birmimgham, B15 2TT, UK \\
$^{2}$ Aix Marseille Univ, CNRS, CNES, Institut Origines, LAM, Marseille, France \\
$^{3}$ Institut d'astrophysique de Paris, UMR 7095 CNRS université pierre et marie curie, 98 bis, boulevard Arago, 75014, Paris, France \\
$^{4}$ Observatoire de Haute-Provence, CNRS, Université d'Aix-Marseille, 04870, Saint-Michel-l'Observatoire, France \\
$^{5}$ Department of Physics and Astronomy, Tufts University, 574 Boston Avenue, Medford, MA 02155 \\
$^{6}$ Astrophysics Group, Keele University, ST5 5BG, UK \\
$^{7}$ Astronomy Unit, Queen Mary University of London, Mile End Road, London E1 4NS, UK \\
$^{8}$ Institute of Astronomy, University of Cambridge, Madingley road, Cambridge CB3 0HA, UK \\
$^{9}$ European Space Agency (ESA), European Space Astronomy Centre (ESAC), Camino Bajo del Castillo s/n, 28692 Villanueva de la Cañada, Madrid, Spain\\ %short version if preferred: European Space Agency, ESAC, Spain
}
\date{Accepted XXX. Received YYY; in original form ZZZ}

\pubyear{2025}

\begin{document}
\label{firstpage}
\pagerange{\pageref{firstpage}--\pageref{lastpage}}
\maketitle

% Abstract of the paper (237 words)
\begin{abstract}Eclipsing binaries are perfect laboratories to measure precise, accurate and model-independent stellar radii and stellar masses, so long as both components are spectroscopically resolved. Resolving both components is difficult in high-contrast binaries, for instance, those composed of an FGK main-sequence star with an M-type companion. In those cases, the secondary can contribute $<1\%$ of the total flux in  optical wavelengths. This makes measuring dynamical masses challenging and has typically only been attempted with large-aperture telescopes (8-10-m). The High-Resolution Cross-Correlation Spectroscopy (HRCCS) method was developed to extract weak emission and transmission spectra for exoplanet atmospheres. This method was  recently adapted and applied to measure dynamical masses in high-contrast binaries. In this work, we apply the HRCCS method to optical spectra of the high-contrast binary and circumbinary planet host Kepler-16~AB, obtained with the SOPHIE spectrograph at the 1.93-m telescope at the Observatoire de Haute-Provence. The secondary, which has a contrast ratio of $\sim 6 \times10^{-3}$, is resolved with a detection significance of 9.5~$\sigma$. We derive dynamical masses with a precision of $1.5\,\%$ and $0.9\,\%$ for the primary and secondary respectively. These are comparable, but slightly higher (within $2-7\,\%$), to previous mass-measurements, which has -within the uncertainties- no implication for the mass of the known circumbinary planet. This work demonstrates that dynamical mass measurements of high-contrast binaries can be done with 2-m class telescopes. We also investigate different analysis protocols to ensure we derive robust uncertainties for dynamical masses.

\end{abstract}


%%%%%%%%%%%%%%%%%%%%%%%%%%%%%%%%%%%%%%%%%%%%%%%%%%

%%%%%%%%%%%%%%%%% BODY OF PAPER %%%%%%%%%%%%%%%%%%

\section{Introduction}

Very low-mass main-sequence stars ($M < 0.35\,{\rm M_{\odot}}$) with precise absolute mass and absolute radius measurements play a crucial role to calibrate stellar evolution models at the bottom of the main-sequence. This incomplete view about the physical properties of these fully-convective stars have far-reaching implications. For instance exoplanets' physical parameters (as well as their atmospheric parameters) directly depend on accurate stellar parameters, which are typically based on stellar evolution models. However, stellar evolution models do not all reproduce observations well. For instance,the `radius inflation problem' \citep[e.g.][]{Casagrande08,Torres10,Spada13,Kesseli18} causes stellar radii from evolutionary models to be under predicted by a few percent. 
This discrepancy is still not fully understood. 

To empirically characterise very low-mass main-sequence stars, a large sample of eclipsing binaries is needed. However few low-mass double-lined eclipsing binaries are known \citep[e.g. DEBCAT; ][]{Southworth2015}. Observations have instead focused on F, G, \& K-type primaries that host eclipsing very low-mass stellar companions. Creating a mass-radius-luminosity-metallicity relation is the goal of the EBLM project \citep[Eclipsing Binaries -- Low Mass; ][]{triaud13,Maxted23}. However, by focusing on eclipsing single-lined binaries (SB1s), the mass and radius of the M-dwarf companion are dependent on accurate estimates for the primary stars mass and radius. Despite highly reliable parameters, this means the EBLM systems are not included in catalogs such as DEBCAT \citep{Southworth2015}, and thus not included to compare stellar models and observations.

EBLM binaries do have an important advantage. Their high-contrast between both stars in optical wavelengths \citep[on average $<0.6\%$,][]{triaud17}  mean (i) that the primary stars can be precisely characterised using standard spectroscopic analysis and stellar evolution models similar to single F, G, \& K-stars \citep{Freckelton24}, and (ii) that  accurate and precise radial-velocities can be obtained to enable the detection of circumbinary exoplanets \citep[e.g.][]{martin19,Standing22}. Detecting a circumbinary planet typically requires dozens of high signal-to-noise spectra, enough that the secondary can be spectroscopically resolved via a careful analysis.

\citet[][hereafter S24]{Sebastian2024a}  recently used the HRCCS method \citep[High-Resolution Cross-Correlation Spectroscopy; ][]{snellen2010,Birkby2018, Sebastian24b} to recover the weak signal of the faint secondary star in the TOI-1338/BEBOP-1 circumbinary system. In this case, optical high-resolution spectra from ESPRESSO at the 8.2-m VLT \citep{pepe21} were used, and precise dynamical masses ($<2\%$) of the binary were obtained. Whilst regularly applied  to detect and characterise exoplanet atmospheres \cite[e.g.][]{Brogi2012,deKok13}, this was the first application of the HRCCS method to measure the absolute masses of a binary system. This method is highly effective at removing the primary's signal from the data (\citetalias{Sebastian24b}). This way the single-lined binary (SB1) was turned into a double-lined (SB2) eclipsing binary. Once combined with new, precise {\it TESS} data \citep{Maxted24} this enabled a full characterisation of this binary system. It also validated the parameters obtained via the traditional SB1 approach normally used by the EBLM project \citep{Davis2024}.

In this paper, we turn our attention to Kepler-16AB, a  $P \sim41\,{\rm d}$  EBLM binary that hosts the first  unambiguously detected circumbinary exoplanet,  Kepler-16\,b \citep[$P \sim229\,{\rm d}$, ][]{Doyle11}. The planet was identified from transits with light curves obtained by the {\it Kepler} mission \citep{Borucki10}.  Kepler-16\, also became the first circumbinary planet with a mass measured with radial velocities \citep{triaud22}. Thanks to the relatively late-type primary (K6), the system has already been measured as a double lined binary by \citet{Bender2012}, who used six spectra, taken with the HRS spectrograph mounted on the 9.2-m HobbyEberly Telescope \citep{tull98}. They applied a standard cross-correlation tool {\tt TODCOR} \citep{Zucker94} to resolve the weak secondary star. The HRS spectra covered the spectral range from $6100 -8870\,${\AA} and the reported accuracies on the dynamical masses were 2.5\,\% on the primary and 1.5\,\% on the secondary respectively.

In this paper, we (i) bench-mark the HRCCS method, using archival data of the well characterised binary system Kepler-16~AB, which were obtained by \cite{triaud22} and (ii) show that we can measure accurately and precisely dynamical masses of both stellar components using optical high-resolution spectra from a 2-m-class telescope.

The data selection and HRCCS analysis are described in Sections~\ref{sec:analysis} and \ref{sec:global_time}. We introduce an improved approach to robustly estimate uncertainties using the HRCCS method in Sec.~\ref{sec:unc} and extract phase- and wavelength depended information of the faint secondary in Sec.\ref{sec:phase_curve}.

\section{SOPHIE data collection}

Between August 2016 and November 2023, 174 spectra have been obtained using the stabilised, fibre-fed high-resolution spectrograph SOPHIE at the 1.93-m telescope of the Observatoire de Haute-Provence in France \citep{Perruchot08}. The SOPHIE data covers a slightly bluer spectral range of $3900 - 6900\,${\AA}, compared to the HRS spectral range \citep{Bender2012}. Hence, the expected contrast between both components ($\sim 0.6$\,\%) is slightly larger.

The data were obtained in the high efficiency mode (R = 40 000), with a typical exposure time of 1800\,s, and an average signal-to-noise-ratio (SNR) of 35 at 550\,nm. The main fibre was placed on the star and the second fibre on the sky to allow optimal subtraction of the sky background. A radial velocity standard star has been observed every night and any possible wavelength drift during each night is monitored using a Fabry-Pérot etalon. All spectra were reduced using the SOPHIE data reduction pipeline \citep[DRS][]{Bouchy09}. The extracted radial velocities from the early part of this data-set have been used in \citet[][hereafter T22]{triaud22} to fit the primary's orbit and to detect the circumbinary planet Kepler-16~b. The data-set we analyse in this paper covers an additional 2.3 years (27 new spectra).

\section{Detrending and CCF analysis}
\label{sec:analysis}

We use one dimensional `s1d' spectra, provided from the SOPHIE DRS. In these data products all échelle orders of the spectrograph are already combined and the spectra are resampled to have a common resolution of 0.01\,\AA. We chose to use the s1d spectra since the combination of orders in the DRS has been optimised to reduce noise at the edges of the detector. For the data preparation we follow closely the methods used in \citetalias{Sebastian24b}. We split the s1d spectra in 60 spectral chunks of 5119 points each and then normalise each spectral chunk -which are aligned in the primary's rest frame- using a polynomial of order 6. For this, we linearly interpolate the spectra using the orbital parameters from \cite{triaud22}. We use this array of aligned spectra also to exclude columns with standard deviation more than 1.9-$\sigma$ higher, compared to the average standard deviation in the normalised chunk. This value has been found experimentally for the data to exclude prominent outliers effectively without affecting the stellar absorption spectrum.

% Outlier removal
Due to the high stability of the SOPHIE spectrograph, the telluric absorption lines in each spectrum should be at constant velocity to within a few m\,s$^{-1}$.  In a first step we identify spectra unusable for the HRCCS processing by analysing the telluric drift of all spectra. We use a telluric line list compiled from the HITRAN database \citep{Gordon2022} including ${\rm H_2O}$, OH, and ${\rm CO_2}$ transitions which were normalised using the Cerro Paranal Sky Model for vacuum wavelengths \citep{noll12,Jones13}. We cross-correlate the SOPHIE orders, which are most affected by telluric absorption lines between $6841$ and $6893${\AA},  with this telluric line list to measure and extract the telluric velocity. We find that most SOPHIE spectra are well calibrated, except for a few outliers. These are the first 12 observations taken before November 2016. The extracted RVs were corrected in \citetalias{triaud22}, but since in this work, we re-calculate the CCF for each spectrum, we opt to not include these spectra in the analysis. We also excluded two observations at BJD = 2458003.32 and 2458398.33\,d, which show an telluric offset of more than 25\,km\,s$^{-1}$. This selection results in 160 SOPHIE spectra, which are used for the HRCCS processing in this work.

\begin{figure}
	\includegraphics[width=\linewidth]{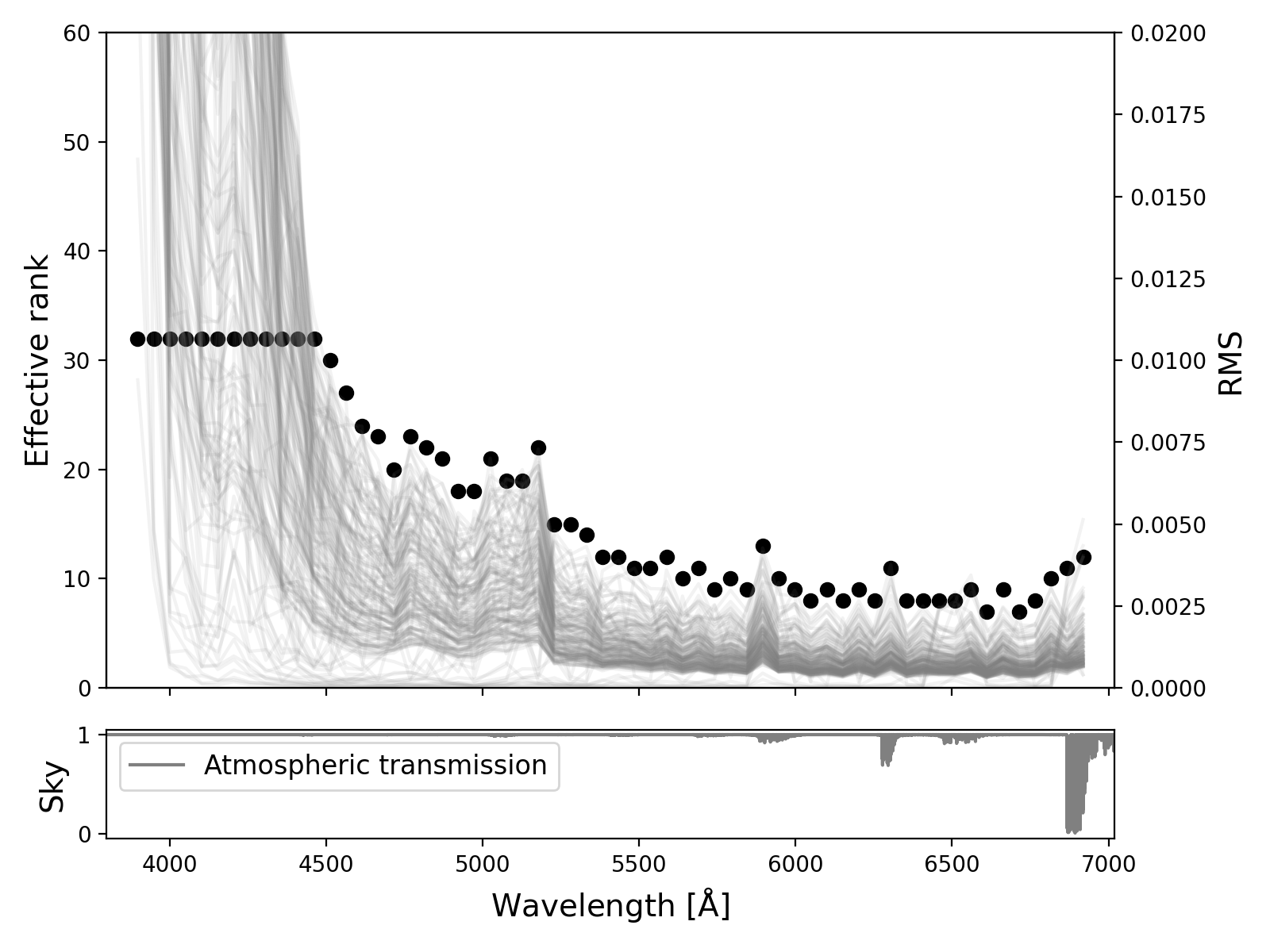}
    \caption{Upper panel: Detrending for SOPHIE data. Black dots: Effective rank, Grey lines, RMS of the residual arrays as a function of wavelength. Lower panel: Atmospheric transmission. Wavelength areas with large RMS in the upper panel, match well with strong telluric lines, which are less correlated in the primary's rest frame.}
    \label{fig:eff_rank}
\end{figure}

% Detrending
In the next step, we remove the absorption spectrum of the primary star. Following \citetalias{Sebastian24b}, we apply a singular-value-decomposition \cite[SVD, ][]{kalman1996}. This is applied directly to each normalised chunk consisting of 160 spectra, which are aligned in the rest frame of the primary star. The SVD decomposes this data into components, which are sorted in descending order according to their correlation in the primary's rest frame. In this way it effectively separates the spectral features of the primary from the data. We apply the `effective rank' \citep{roy07} to select the optimum number of SVD components for each chunk. In order to handle noisy chunks, we set a maximum number of removed SVD components to 32, which is a fifth of the total number of components (160). The number of SVD components, removed in each chunks is shown in Fig~\ref{fig:eff_rank}. The median number of components is 12.5, which is less than 8\,\% of this total number (reduced rank, see \citetalias{Sebastian24b}). With an average contrast of 0.6~per-cent, the SNR of the secondary star, per resolution element, in each SOPHIE spectrum, is on the order of 0.02. According to Table~4 in \citetalias{Sebastian24b}, we thus expect that this SVD detrending has removed less than 5~\% of the secondary's spectral features, and thus, has kept its signal mainly intact.  

% Saltire
To detect the signal of the secondary in the detrended data, we first derive a CCF with a line mask, optimised for M-dwarfs. Second, we make use of the K-focusing process \citep{Sebastian2024a}, which enhances the secondary's weak CCF by combining it in the secondary's rest frame. Since the rest velocity $V_{\rm rest,2}$, as well as the semi-amplitude $K_{2}$ of the secondary are a priori unknown, this combination is explored in the $ K_{2} - V_{\rm rest}$ plane. The remaining orbital parameters which are the period ($P$), time of periastron ($T_{\rm 0,peri}$), eccentricity ($e$), and the argument of periastron ($\omega$) are adopted from \citetalias{triaud22} and kept fixed in this analysis. For the CCF, we use a line mask for an M5 star, commonly used to extract precise RVs within the data reduction pipeline of the ESPRESSO spectrograph at the ESO-VLT\footnote{ESO pipelines are publicly available on \hyperlink{https://www.eso.org/sci/software/pipe_aem_main.html}{eso.org}} \citep{pepe21}. The rest velocity of the primary has been measured to be $V_{\rm rest,1} = -33.8\,{\rm km\,s^{-1}}$ by \citetalias{triaud22}. We therefore derive CCFs in a grid from -60 to 0\,km\,s$^{-1}$, with a step width corresponding to the SOPHIE pixel resolution (1.5\,km\,s$^{-1}$)\footnote{\href{http://www.obs-hp.fr/guide/sophie/thorium_psf.html}{SOPHIE instrument PSF}}. We cover $ K_{2}$ in a grid from 25 to 75\,km\,s$^{-1}$, again in a step width of 1.5\,km\,s$^{-1}$. Before we combine all CCFs we exclude data at orbital phases where the velocity of the secondary's absorption lines differs less than 10\,km\,s$^{-1}$ from the primary's absorption lines. In this way we exclude 19 spectra, making sure any remaining artefacts from the primary in the detrended data do not interfere with the radial velocity analysis.

\begin{figure}
	\includegraphics[width=\linewidth]{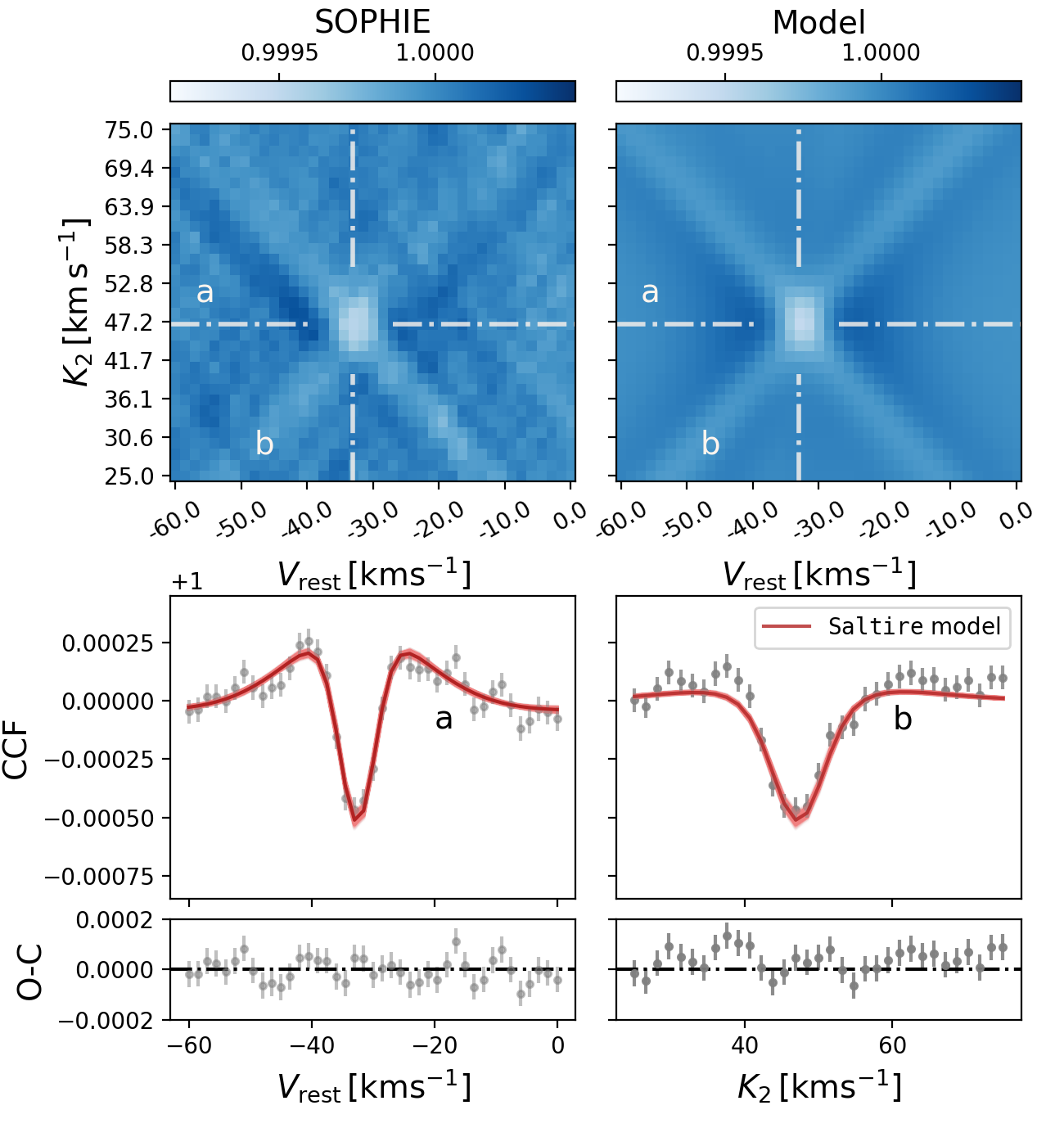}
    \caption{Upper panel: Cross-correlation map and {\tt Saltire} model for SOPHIE data. Black dotted lines show the positions of slices (a,b), displayed in the lower panel. Lower panel: Slice through CCF map at maximum significance. Red line: best-fitting {\tt Saltire} model, red shaded area: $1\sigma$ Uncertainties from the MCMC. Error bars represent the MCMC jitter term ($\sigma_{\rm jit}$) from the two dimensional fit.}
    \label{fig:SOPHIE_saltirefit}
\end{figure}

Figure~\ref{fig:SOPHIE_saltirefit} shows the resulting cross-correlation map in the $ K_{2} - V_{\rm rest}$ plane. This map is fitted using the {\tt Saltire} model \citep{Sebastian2024a} to derive the parameters $ K_{2}$ and $V_{\rm rest}$. The model fits the CCF using a K-focused double Gaussian function. This function models symmetrical side-lobes to the main CCF with five parameters. These are the mean height ($h$) of the CCF outside the signal, the standard deviations ($\sigma_1$ and $\sigma_2$) of two Gaussian functions with intensities $A_1,\,A_2$, the quotient $A_2/A_1$, as well as the sum $A_1+A_2$ of both intensities. Similar to \citetalias{Sebastian24b} we also sample the posterior probability distribution, using the Markov chain Monte Carlo (MCMC) code \texttt{emcee} \citep{Foreman-Mackey13}, which is implemented in the {\tt Saltire} model. The signal of the secondary is well detected with a detection significance of 9.5~$\sigma$. Fit parameters are presented in Table~\ref{tab:measures}.
 
\section{Global time fitting}
\label{sec:global_time}

\begin{figure*}
	\includegraphics[width=\textwidth]{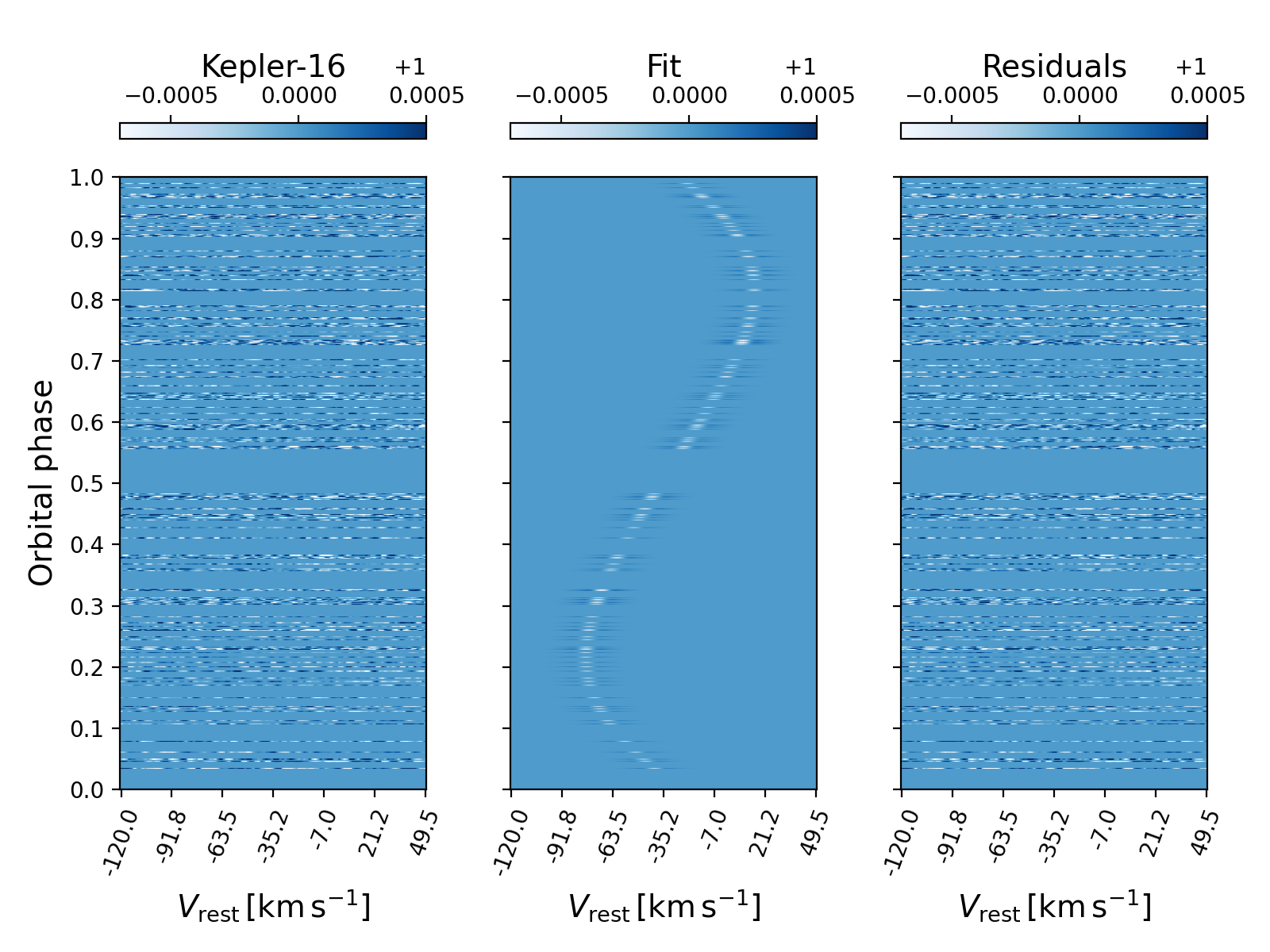}
    \caption{Global time fitting of CCF data of Kepler-16~B. Left panel: CCF data from detrended SOPHIE spectra, folded to the orbital phase of the binary. Middle panel: Best fit double Gaussian model. Right panel: Residuals. Phases without data are presented as blank blue sections.}% Only CCF sections close to the trail are used for the fit. }
    \label{fig:glob_fit1}
\end{figure*}

The main advantage of the K-focusing method is the possibility to combine and detect very weak signals, such as planet atmospheres or very faint M-dwarfs. The drawback of this method is that the data in such CCF maps are highly correlated, making it difficult to constrain uncertainties \citep{Sebastian2024a}. In the case of Kepler-16~B, we can use the relatively high detection significance to resolve the CCF trail directly in the time frame. For each detrended spectrum, we first correct for the barycentric drift and compute a CCF with the same line mask for a M5 star -we used before- in a grid from -120 to 50\,km\,s$^{-1}$, with a step width of 1.5\,km\,s$^{-1}$. This ensures the secondary's signal is covered in all orbital phases, taking the measured rest velocity and semi-amplitude into account. Similar to Sec.~\ref{sec:analysis} we exclude 19 of the 160 spectra at orbital phases where the velocity of the secondary's absorption lines differs less than 10\,km\,s$^{-1}$ from the primary's absorption lines. The left panel of Figure~\ref{fig:glob_fit1} shows the resulting CCF functions as a function of the binary's orbital phase. The CCF trail of Kepler-16~B is clearly resolved for a large part of the orbit. The CCF trail is relatively weak. The measured CCF contrast is smaller than the RMS noise of the CCFs ($\rm SNR_{trail} = 0.8$). Therefore, the CCF profile is hardly distinguishable from noise in individual CCFs.

Different to classical CCF fitting, which fits a Gaussian function to each individual CCF, we need to apply a global fit simultaneously to all CCF data, using a forward model of the CCF trail. All CCFs are simultaneously fitted to a double Gaussian function consisting of a main CCF signal as well as a side-lobe component, as used in the {\tt Saltire} model. The position of the CCF signal in the CCF trail model is defined by its Keplerian movement following
\begin{equation} \label{reflex}
    V_{\rm r,2} = K_{2} [\cos(\nu + \omega_{2}) + e \cos(\omega_{2})],
\end{equation}
where $K_{2}$ is the secondary's semi-amplitude, and $\nu$ the true anomaly at the time of mid-exposure, which is determined from the orbital period $P$ and the time of periastron $T_{\rm 0,peri}$. Both were precisely measured by \cite{triaud22}, who also obtained parameters for the eccentricity ($e$) and the argument of periastron of the primary ($\omega$). The argument of periastron of the secondary ($\omega_{2}$) is derived from the primary by $\omega_{2}=\omega - \pi$. Different from the K-focusing method these Keplerian parameters are no longer kept fixed, and are now used as free parameters.

We sample the posterior probability distribution, of all free parameters using \texttt{emcee}. Given the low signal to noise of the CCF trail, we use tight uniform priors for the four Keplerian parameters $P$, $T_{\rm 0,peri}$, $e$, and $\omega_{2}$ using the uncertainties measured by \citetalias{triaud22}. We use the fit of the {\tt Saltire} model to obtain starting parameters, for $K_{2}$, $V_{\rm rest}$, and the double Gaussian function which are listed in Table\,\ref{tab:measures}. These are sampled with wide uniform priors over 6\,000 calls for 42 parallel chains, rejecting the burn-in samples (the first 1,000 samples) of each walker and thin the remaining samples by a factor 5. The resulting parameters and uncertainties are listed in Table \,\ref{tab:measures}. The middle panel of Figure~\ref{fig:glob_fit1} shows the best fitting model of CCF functions and the right panel the residual CCF's. The relatively low SNR is reflected in a jitter term ($\sigma_{\rm jit}$) which is about ten times higher compared to the K-focusing method. We find comparable parameters for the mean CCF shape as well as comparable uncertainties for the orbital parameters $K_{2}$ and $V_{\rm rest}$ to the {\tt Saltire} fitting. Before we compare the measured parameters, we analyse our results for systemic uncertainties in the following section.

%Add table for saltire fit and global fit results!!!
\begin{table*}
	\centering
	\caption{Priors and measured parameters from {\tt Saltire} and global time MCMC fitting for detrended SOPHIE data. Errors of individual parameters are mean systematic uncertainties from time splitting, (Sec.~\ref{sec:time_splitting})). Fit uncertainties are given in brackets for comparison.}
	\label{tab:measures}
	\begin{tabular}{llll}
    \hline
    \multicolumn{4}{c}{Fixed parameters and priors} \\
    Parameters & Values & {\tt Saltire} & Global fit\\\hline
    $P [{\rm d}]$ & $41.077772$ & fixed & $\pm0.000051^{\ast}$ \\
    $T_{\rm 0,peri} [{\rm BJD}]$ &  $2\,457\,573.0984$ & fixed & $\pm0.0047^{\ast}$ \\
    $e$ & $0.15994$ & fixed & $\pm0.00010^{\ast}$ \\
    $\omega [{\rm rad}]$ & $4.60194$ & fixed & $\pm0.00051^{\ast}$ \\\hline
    %$i [{\rm ^{\circ}}]$ & $89.658\pm0.146^{\dagger}$\\\hline
    \multicolumn{4}{c}{Measured parameters}  \\
    Parameters & & {\tt Saltire} & Global fit\\\hline
    $P [{\rm d}]$ && -- & $41.077777\pm0.000035$ \\
    $T_{\rm 0,peri} [{\rm BJD}]$ && -- & $2\,457\,573.0995\pm0.0032$ \\  
    $e$ && -- & $0.15962\pm0.00037$ \\ 
    $\omega [{\rm rad}]$ && -- & $4.60184\pm0.00034$ \\ 
    $K_{2} [{\rm km\,s^{-1}}]$ && $47.16\pm0.29\,(0.08)$ & $46.88\pm\,0.28$ (0.21) \\
    $V_{\rm rest,2} [{\rm km\,s^{-1}}]$ && $-32.56\pm0.30\,(0.06)$ & $-32.65\pm0.22$ (0.15) \\
    $A_{1} + A_{2}$   && $-(4.81\pm0.61\,(0.16)\times10^{-4}$ & $-(5.42\pm0.48\,(0.36))\times10^{-4}$ \\
    $A_{2} / A_{1}$   && $-0.421\pm0.015\,(0.017)$ & $-0.363\pm0.062\,(0.057)$ \\
    $\sigma_1\,[\rm km\,s^{-1}]$ && $2.93\pm0.23\,(0.23)$ & $2.85\pm0.18\,(0.22)$ \\
    $\sigma_2\,[\rm km\,s^{-1}]$ && $10.82\pm2.6\,(0.58)$ & $8.9\pm5.2\,(1.2)$ \\
    %$h$                   && $0.999958\pm0.000006$ & $0.99995\pm0.00004$ \\
    $\sigma_{\rm jit}$    && $(5.1\pm0.1)\times10^{-5}$ & $(65.2\pm0.4)\times10^{-5}$ \\

    \hline
    $^{\ast}$ From \cite{triaud22}\\
    %\multicolumn{4}{l}{$^{\dagger}$ `Best' parameters from \cite{Bender2012}}  \\
    
    \end{tabular}
\end{table*}

\section{Uncertainties from partial data}
\label{sec:unc}

Different to the fitting of CCF maps in $K_{\rm p}- V_{\rm rest}$ plane from the K-focusing method, we expect the uncertainties derived from our global time fitting to be unaffected from correlated noise. We use the large dataset of 141 (160-19) SOPHIE spectra to analyse the noise properties of our dataset in two different methods and compare them with the uncertainties, derived from our MCMC fit. We employ two different methods to test our uncertainties, and ensure they are robust. When measuring absolute masses at the per-cent level, this exercise is an important step that ensures  reliability.

\subsection{Time splitting analysis}
\label{sec:time_splitting}

 Similar to \citetalias{Sebastian24b}, we, thus, evaluate the uncertainties of the secondary fit by splitting the dataset of 141 spectra in time into $n=4, 6,\rm{ and}\,8$ sets of 35, 23, and 18 spectra respectively. We measure the parameters of each set (i) by fitting in the $K_{\rm 2}- V_{\rm rest}$ plane using the {\tt Saltire} code and (ii) by performing the global time fitting, introduced in Sec.~\ref{sec:global_time}. Each set is analysed individually and we estimate the systematic uncertainties as the standard error of the parameters of each individual set. Due to the low number of sets, we have to take the Bessel's correction into account. The fractional uncertainty of deriving the standard error for such few sets ($n$) can be estimated as $1/\sqrt{2n-2}$ \citep{Topping1972}. As presented in Table~\ref{tab:time_split}, the fractional uncertainties are larger, when splitting the data in fewer sets. We thus, have to take them into account when comparing the results for different numbers of sets.
 
 The results for both fitting methods are presented in Table~\ref{tab:time_split}. We find that the systematic uncertainties are well in agreement (i) for the two different fitting methods and (ii) for a different number of sets, when taking the fractional uncertainties into account. We finally derive a mean systematic uncertainty ($\overline{X}_{\rm w}$) from the different sets, which we weight by the inverse of the fractional uncertainties. 
 
 We find that this mean systematic uncertainty is only slightly larger (33\,\%), compared to the fit uncertainty from the global time fitting (0.21\,km\,s$^{-1}$ , see Sec.~\ref{sec:global_time}). Taking the fractional uncertainty into account this a $1~\sigma$ agreement. In contrast, this mean systematic uncertainty it is about 4 times larger, compared to the fit uncertainty using  {\tt Saltire}. This means that, the global time fitting can be used to robustly estimate the uncertainties of the orbital parameters.

 We conservatively use the mean systematic uncertainties of each parameter from the time splitting to estimate the uncertainties from the {\tt Saltire} and global time fitting, as presented in Table~\ref{tab:measures}. In brackets we add the fit uncertainties from the MCMC for comparison. Taking these uncertainties into account, we find that the semi-amplitude of the secondary is within a 1~$\sigma$ agreement between the K-focusing and the global time fitting methods.  
 
\begin{table}
	\centering
	\caption{Systematic Uncertainties of the M-dwarfs semi-amplitude ($K_2$) derived from time-splitting of the SOPHIE data in different number of sets. Right column: The statistical factional uncertainty of the derived systematic uncertainty, due to the low number of sets. The mean systematic uncertainty ($\overline{X}_{\rm w}$), is weighted by this fractional uncertainties.}
	\label{tab:time_split}
	\begin{tabular}{lccr} % four columns, alignment for each
		\hline
		Number of sets & {\tt Saltire} fit & Global time fit & Fractional uncertainty\\
                         & [km\,s$^{-1}$] & [km\,s$^{-1}$] & [\%]\\
		\hline
		4 & 0.23 & 0.21 & 41\\
		6 & 0.27 & 0.31 & 32\\
		8 & 0.35 & 0.30 & 27\\\hline
      $\overline{X}_{\rm w}$ & 0.29 & 0.28 & \\
		\hline
	\end{tabular}
\end{table}

\subsection{Bootstrapping analysis}

 Instead of splitting the sample as a function of time, we can also analyse randomly selected samples with a constant number of spectra. This method has been successfully applied to validate the measurement of exoplanet atmospheres, which are affected from correlated noise \citep[e.g.][]{Hoeijmakers2020}. In this study, we select partial samples representing 50\,\% of the $m=141$ spectra.
 
 We draw 200 different partial samples arbitrarily, following the condition that partial samples of size $m$ can only contain a certain amount of $s_{\rm max}$ similar spectra to previous partial samples, with $s_{\rm max} = 0.5 \times ({\rm percentage}/100) \times m$  are allowed. With this, we ensure that individual drawn samples are different from each other. That means a drawn partial sample is not selected if it contain more than 17 spectra, which were drawn together in any previous partial sample.  We derive the bootstrapping uncertainty for $K_{\rm p}$ as the standard deviation of the fit results, for all 200 partial samples and correct for the reduced number of spectra by multiplying with $\sqrt{2}$.
 We finally repeat this experiment $3 \times$, which results in an average uncertainty of 0.24\,km\,s$^{-1}$ on $K_2$. 
 
This uncertainty from 50\,\% bootstrapping is consistent with both the fit uncertainty from the global time fit and with the systematic fit uncertainty (Sec.~\ref{sec:time_splitting}). We therefore find that the bootstrapping is a good metric to estimate the uncertainties of noisy data (faint signals), which cannot be separated in several parts. Since we have shown in Sec.~\ref{sec:time_splitting} that the global time fitting as well as fitting in the $K_{\rm 2}- V_{\rm rest}$ plane (K-focusing) result in similar results for partial samples, the bootstrapping approach can also be applied to derive uncertainties for very noisy data, where the K-focusing is the only option.

\section{Dynamical masses}

Combining the high precision measurement of the primary's semi-amplitude from \citetalias{triaud22} and the semi-amplitude of the secondary measured in this work, we can derive dynamical masses for both stars. Here, we use the inclination, measured from high-precision {\it Kepler} light-curves, which is $ i = 90.3401\pm0.0019\,\rm ^{\circ}$ \citep{Doyle11}. We follow the IAU recommended equations \citep[Table 3 in][]{2016AJ....152...41P} to derive the dynamical masses ($M_{1},M_{2}$), the mass ratio ($q 
 = M_{2}/M_{1}$), and semi-major ($a$) axis directly from the measured parameters as listed in Table~\ref{tab:measures}.

The derived parameters are listed in Table~\ref{tab:bin_par} together with dynamical masses derived by \cite{Bender2012} and the photometric-dynamical model by \cite{Doyle11}. We find that the derived dynamical masses from SOPHIE data are on average $\sim 3\,\sigma$ larger, compared to the literature, causing the binary system to be about 7\,\% more massive. This discrepancy could be caused by the low number of only six spectra, used by \citet{Bender2012}. Comparing to the photometric-dynamical literature masses, the binary system appears to be only 2\,\% more massive, with the masses of both stars being compatible within $\sim 1.5\,\sigma$. This slight increase in mass would also affect the mass of the detected exoplanet. Given the reported uncertainties of the planet's radial velocity by \citetalias{triaud22}, the planet's mass is still in perfect agreement to the reported mass, which is why we do not derive an updated planet mass in this work.

\begin{table}
	\centering
	\caption{Comparison of Masses for Kepler-16 AB.}
	\label{tab:bin_par}
	\begin{tabular}{ll}
    \hline
    \multicolumn{2}{c}{Photometric-dynamical model}  \\
    parameter & value\\\hline
    $q$ & $0.2937\pm0.0006^{\dagger}$\\
    $M_{1} [{\rm M_{\sun}}]$ & $0.6897\pm0.0035^{\mathsection}$ \\
    $M_{2} [{\rm M_{\sun}}]$ & $0.20255\pm0.00066^{\mathsection}$ \\
    $a [{\rm AU}]$ & $0.22431\pm0.00005^{\mathsection}$\\
    \hline
    
    \multicolumn{2}{c}{Dynamical masses from literature}  \\
    parameter & value\\\hline
    $K_{1} [{\rm km\,s^{-1}}]$ & $13.642\pm0.010^{\dagger}$\\
    $K_{2} [{\rm km\,s^{-1}}]$ & $45.56\pm0.47^{\dagger}$\\
    $q$ & $0.2994\pm0.0031^{\dagger}$\\
    $M_{1} [{\rm M_{\sun}}]$ & $0.654\pm0.017^{\dagger}$ (2.6\,\%)\\
    $M_{2} [{\rm M_{\sun}}]$ & $0.1959\pm0.0031^{\dagger}$ (1.6\,\%)\\\hline
    
    \multicolumn{2}{c}{Dynamical masses from SOPHIE data}  \\
    parameter & value\\\hline
    
    $K_{1} [{\rm km\,s^{-1}}]$ & $13.6787\pm0.0015^{\ast}$\\
    $K_{2} [{\rm km\,s^{-1}}]$ & $46.88\pm0.28$ \\
    $q$ & $0.2918\pm0.0017$\\
    $M_{1} [{\rm M_{\sun}}]$ & $0.704\pm0.011$ (1.5\,\%)\\
    $M_{2} [{\rm M_{\sun}}]$ & $0.2054\pm0.0019$ (0.9\,\%)\\
    $a [{\rm AU}]$ & $0.2257\pm0.0010$\\

    \hline
    
    \multicolumn{2}{l}{$^{\mathsection}$ From \cite{Doyle11}}  \\
    \multicolumn{2}{l}{$^{\dagger}$ From \cite{Bender2012}}  \\
    $^{\ast}$ From \cite{triaud22}\\
    \end{tabular}
    \\
\end{table} 

\section{Phase-curve and low-resolution spectrum}
\label{sec:phase_curve}

The detrended data allow us to actually extract the spectrum of the M-dwarf. This can be used to understand any phase-dependent intensity variations, or to identify spectral features. 

\begin{figure*}
	
        \includegraphics[width=\linewidth]{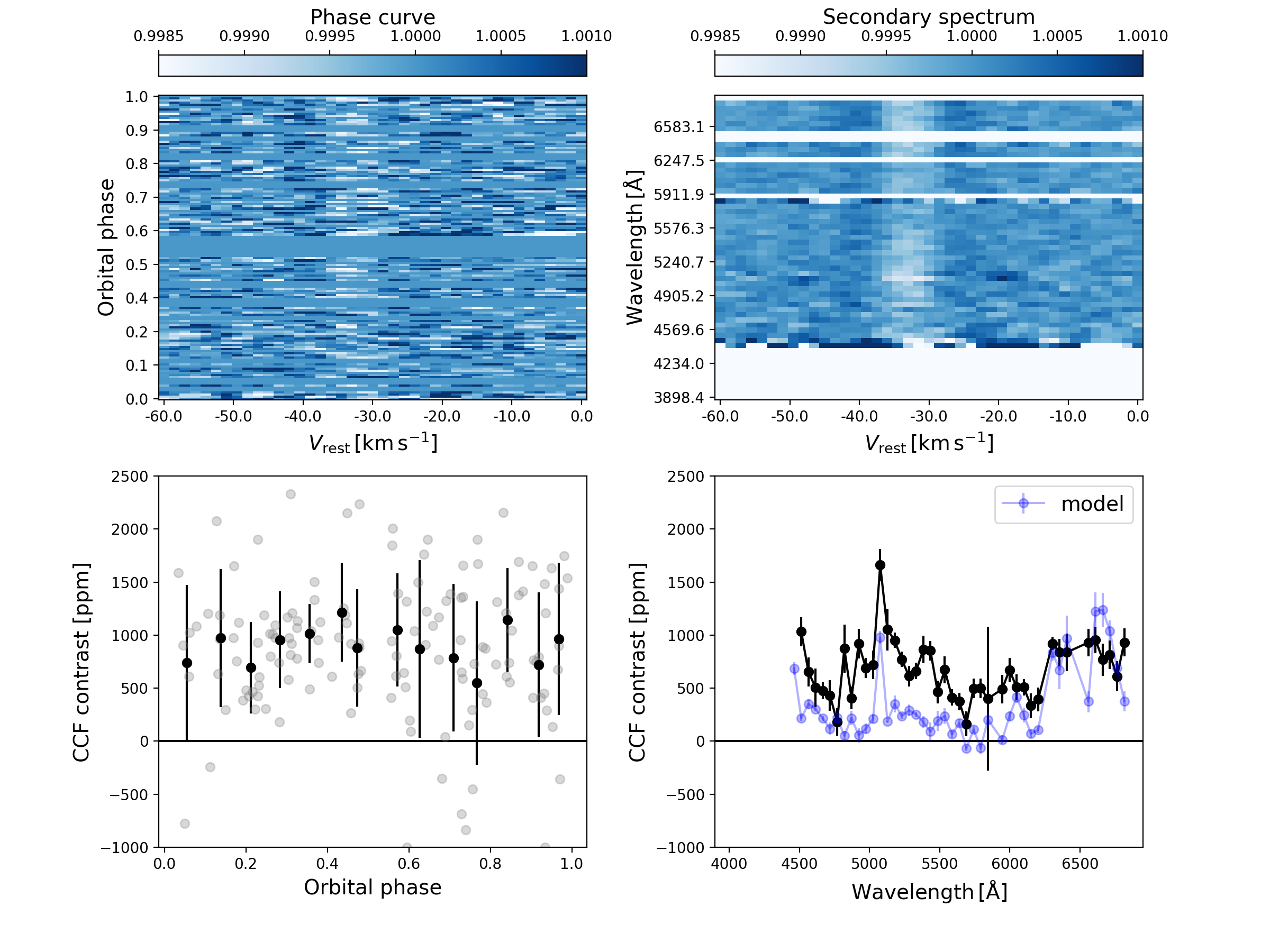}
    \caption{Upper panels, CCF functions in the secondary's rest-frame. Left: as a function of the orbital phase, Right: as a function of the Wavelength. Lower panels, measurements of individual CCF contrasts (grey dots) and binned data (black dots). Left: Phase curve of the M-dwarf secondary, Right: residual flux as a function of wavelength, showing a low resolution spectrum of the M-dwarf (Black dots) and a residual spectrum of the secondary, derived from a PHOENIX model spectrum (Blue dots).}
    \label{fig:phase curve}
\end{figure*}

We can extract the phase dependent CCF contrast from the CCF trail directly. The CCF contrast is a measure of the secondary contribution to the total light. Following \citetalias{Sebastian24b} we derive the CCF trail in the rest frame of the secondary and fit each CCF in the time frame with a Gaussian function. The left panels of Figure~\ref{fig:phase curve} show the extracted CCF contrast. Within the uncertainties the secondary's signal appears to be constant over the orbital phase, which is expected for a low mass star at visual wavelengths (for an exoplanet we would expect a day-night variation).

We combine all 141 spectra in the rest frame of the secondary and measure the combined CCF contrast for each of the 60 spectral chunks individually. In this way, we can extract the wavelength-resolved signal of the secondary, thus a low-resolution residual spectrum of the M-dwarf. The right panels of Figure~\ref{fig:phase curve} show the residual spectrum. Since the CCF contrast is a function of the number of absorption lines and used line mask, the features do not necessarily correspond to individual spectral features, as seen in low-resolution spectra. This extracted spectrum can be compared to observed or to modelled spectra, to confirm its origin from the M-dwarf secondary. 

To attempt this, we use a simulated PHOENIX model spectra \citep{Husser13}. Stellar parameters of the primary and the radius of the secondary were published by \cite{Doyle11}. For the secondary, we use typical parameters for main-sequence stars \citep{PM2013} to estimate the stellar effective temperature. We finally use PHOENIX model spectra with $T_{\rm eff,2}=4400\,{\rm K}$ and $\log\,g_\star=4.5$ for the primary and $T_{\rm eff,2}=3100\,{\rm K}$ and $\log\,g_\star=5.0$ for the secondary, assuming a solar metallicity ($\rm [Fe/H]=0.0$).
Following \citetalias{Sebastian24b} we correct the model wavelengths from vacuum to air, following \cite{Morton00}, then we match the spectral resolution to SOPHIE ($R\sim40\,000$) by convolution with a Gaussian kernel and correct the continuum for the radii of both stars, as measured by \cite{Doyle11} from precise {\it Kepler} light-curve modelling ($R_1=0.6489\,R_{\rm \sun}$ and $R_2=0.22623\,R_{\rm \sun}$). For the final model, we used the secondary spectrum, corrected for the continuum of the primary. This noiseless model spectrum is separated into 60 spectral chunks, normalised, and cross-correlated in the secondary rest-frame to reproduce how the observations are treated. The measured CCF contrast of this model is shown in the right lower panel of Figure~\ref{fig:phase curve}. While the average CCF contrast of the model is smaller, compared to the SOPHIE data, dominant features -such as the increase of CCF contrast at 6300\,\AA- are present in the model as well.

\section{Conclusions}

Kepler-16 AB, is an eclipsing binary hosting the first circumbinary planet, which is detected with both high-precision light-curve \citep{Doyle11} and radial velocity measurements \citep{triaud22}. Intensive follow-up in the recent years made it one of the most studied circumbinary systems known to date. The faint M-dwarf secondary makes it difficult to measure the radial velocity of the secondary component and thus model-independent dynamical masses. This has only been possible using a 9-m class telescope \citep{Bender2012} using standard cross-correlation methods.

In this study, we use HRCCS (High-Resolution Cross-Correlation Spectroscopy), commonly used to detect and analyse exoplanet atmospheres. This method has recently been qualified to measure precise dynamical masses of high-contrast binaries \citep{Sebastian24b}. The power of this method is the use of a singular value decomposition (SVD), which can be used to remove the atmospheric lines of the primary component completely from the data, but keeps more than 95\,\% of the secondary's atmospheric lines intact. We apply HRCCS to 160 SOPHIE spectra to detect and measure the semi-amplitude of the faint secondary star, which is the first application of this kind to data of a 2-m class telescope. We use two different fitting methods to analyse the cross-correlation functions (CCF's) of our data. The `K-focusing', which is fitted using the {\tt Saltire} model \citep{Sebastian2024a} and a global time fit, introduced in this work. We find that both fitting methods result in comparable results. 

We furthermore analyse our measurements for systematic uncertainties, by fitting partial SOPHIE data (i) by manually splitting the data in different sets and (ii) by using a bootstrap algorithm. We find that the uncertainties, derived from the global time fitting are well in agreement to the systematic uncertainties. We therefore conclude, that the global fit is superior to the `K-focusing' method, but requires a sufficiently strong CCF contrast to actually fit the CCF in the time domain ($\rm SNR_{trail} = 0.8$ for Kepler-16~B in SOPHIE data). We furthermore find that the bootstrapping allows to estimate accurate uncertainties for those low SNR data which can only be fitted using the `K-focusing' method.

As listed in Table~\ref{tab:bin_par}, we derive precise dynamical masses of Kepler-16 AB with uncertainties of 1.5\,\%, and 0.9\,\% for the primary and secondary respectively. We furthermore find that the dynamical masses, derived with HRCCS, are slightly ($2-7\,\%$) higher for both binary components. This is well in agreement ($\sim 1.5\,\sigma$) to the photometric-dynamical model \citep{Doyle11}. For the dynamical masses presented in \cite{Bender2012}, we only find a $\sim 3\,\sigma$ agreement, which might be caused by the low number of only six measurements, which have been used in that study. Nevertheless, the derived masses have, within the uncertainties, no implications to the mass of the circumbinary planet published by \citep{triaud22}. 

We have thus successfully confirmed the application of the HRCCS method to SOPHIE data and pushed down its application for a high-contrast binaries using an optical high-precision spectrograph on a 2-m class telescope. Thanks to the relatively late spectral type of Kepler-16~A (K6), the contrast ratio allowed to measure dynamical masses for this $0.2\,M_{\rm \sun}$ M-dwarf. We expect that the HRCCS method will be applicable to measure precise secondary masses of $ \sim 0.35\,M_{\rm \sun}$ and above for high-contrast EBLM binaries using the SOPHIE spectrograph.

\section*{Acknowledgements}

We thank the French National Programme in Planetology (PNP) for awarding the telescope time on the 193cm telescope at Observatoire de Haute-Provence (OHP) that led to this paper. Time was granted under proposal IDs: 2024A\_PNPS9, 2023B\_PNPS3, 2023A\_PNPS7, 2022B\_PNPS2, 2022A\_PNPS7, 2018B\_PNPS3, 2019A\_PNP001, 2018B\_PNP001, 2018A\_PNP002, 2017B\_PNP001 \& 2017A\_PNP003 (PI Santerne) as well as the DDT programme DDT\_2016A (PI Triaud). We also thank the kind staff at OHP, particularly the night time operators.

This research is funded from the European Research Council (ERC) through the European Union’s Horizon 2020 research and innovation programme (grant agreement n$^\circ$803193/BEBOP), and from the ERC/UKRI Frontier Research Guarantee programme (grant agreement EP/Z000327/1/CandY). We received funding from the French Programme National de
Physique Stellaire (PNPS) and the "Programme National de Planétologie" (PNP) of CNRS/INSU co-funded by CNES.
MRS acknowledges support from the European Space Agency as an ESA Research Fellow.
%%%%%%%%%%%%%%%%%%%%%%%%%%%%%%%%%%%%%%%%%%%%%%%%%%
\section*{Data Availability}

The reduced spectra are available at the SOPHIE
archive: \url{http://atlas.obs-hp.fr/sophie/}. Data obtained under Prog.IDs 16.DISC.TRIA, 16B.PNP.HEBR, 17A.PNP.SANT, 17B.PNP.SANT, 18A.PNP.SANT, 18B.PNP.SANT, and 19A.PNP.SANT.



\bibliographystyle{mnras}
\bibliography{library}% if your bibtex file is called example.bib

%\begin{thebibliography}{99}
%\end{thebibliography}

%%%%%%%%%%%%%%%%%%%%%%%%%%%%%%%%%%%%%%%%%%%%%%%%%%

%%%%%%%%%%%%%%%%% APPENDICES %%%%%%%%%%%%%%%%%%%%%

\appendix

\section{Radial velocity drift of the primary}

\begin{figure}
	
        \includegraphics[width=\linewidth]{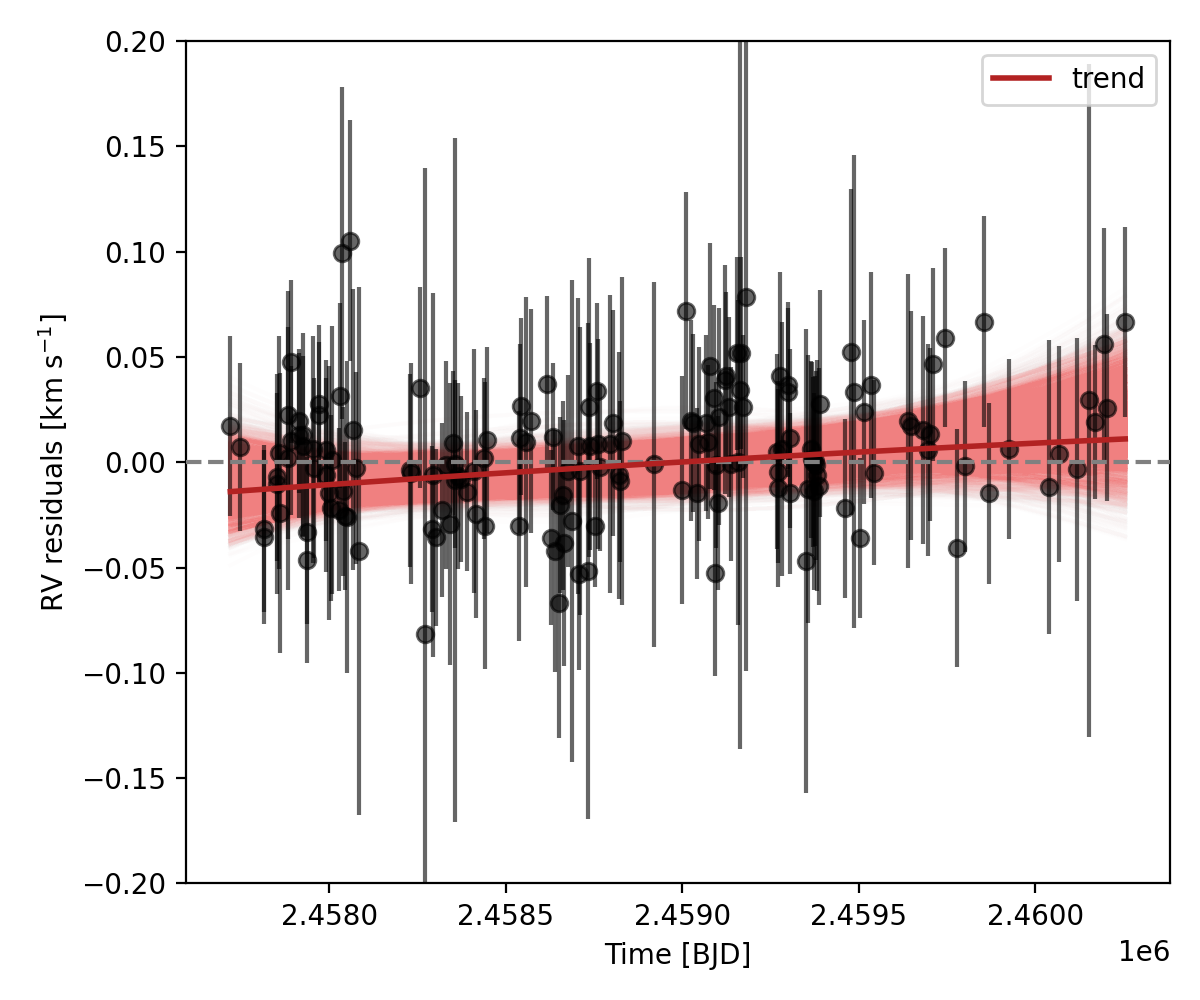}
    \caption{Trend from global time fitting of the primary. Red line: Best fit of the trend, using a cubic drift. Shaded area: $1\,\sigma$ uncertainties from the MCMC fit. Black dots: extracted RV residuals for clarity (not used for the fit).}
    \label{fig:trend}
\end{figure}

 In the precise analysis of \citetalias{triaud22} not only the planet, but also a cubic drift on the order of $\sim 20\,{\rm m \,s^{-1}}$ has been reported in the reflex motion of the binary, which is interpreted as part of the magnetic cycle of the primary star. We, thus, explore the reported drift of the primary. In this, we cross-correlate the 160 SOHPIE spectra with a line mask for an K5 star, which is typically used to extract precise RVs within the data reduction pipeline of the HARPS spectrograph \citep{Mayor03} at the ESO-3.6-m telescope. We then use our global time fitting to precisely measure the primary's orbit parameters. Here we apply an MCMC with wide uniform priors for all orbital parameters and add a cubic drift with two additional parameters. Figure~\ref{fig:trend} shows the best fitting cubic trend as well as the $1\,\sigma$ uncertainties to the systemic velocity and cubic drift. We fit our double Gaussian function to each spectrum to extract the individual RVs. These RVs are less precise, compared to the data products from the SOPHIE pipeline, which is why we don't use them for the fit. We only use them for clarity in Figure~\ref{fig:trend}, after the binary and planet orbits as reported in \citetalias{triaud22} have been removed. These residual RVs show clearly that the drift -also reported by \citetalias{triaud22}- does not continue with the new data (after BJD = 2459459.36\,d). Furthermore, the drift we measure from our global time fit is within the uncertainties not significantly different from zero. This supports the hypothesis by \citetalias{triaud22} that the initial drift is an artifact from the magnetic cycle (1900 to 2400 days) which is well covered by seven years of data, used in this study. As a consequence we fix the drift to zero in all measurements, presented in this study.   

 %2459459.3589

%%%%%%%%%%%%%%%%%%%%%%%%%%%%%%%%%%%%%%%%%%%%%%%%%%

% Don't change these lines
\bsp	% typesetting comment
\label{lastpage}
\end{document}